\documentclass[authoryear,preprint,12pt]{elsarticle_WP}

\usepackage{amssymb}

\usepackage[utf8]{inputenc}
\usepackage{amsthm}
\usepackage{amsmath}
\usepackage{amssymb}
\usepackage{amsfonts}
\usepackage{algorithm}
\usepackage{algorithmicx}
\usepackage{algpseudocode}
\usepackage{comment}
\usepackage{natbib}
\usepackage{hyperref}
\usepackage{xcolor}
\usepackage{setspace}
\usepackage{geometry}
\usepackage{multirow}
\usepackage{subcaption}

\renewcommand{\det}{\textrm{det}}
\newcommand{\intd}{\textrm{d}}
\newcommand{\mvec}{\boldsymbol{m}}
\newcommand{\mvecc}{\tilde{\boldsymbol{m}}}

\newcommand{\vvec}{\vec{\boldsymbol{v}}}
\newcommand{\xvec}{\boldsymbol{x}}
\newcommand{\yvec}{\boldsymbol{y}}
\newcommand{\zvec}{\boldsymbol{z}}
\newcommand{\Xvec}{\boldsymbol{X}}

\newcommand{\Zvec}{\boldsymbol{Z}}

\journal{Spatial Statistics}

\begin{document}

\begin{frontmatter}



\title{Spherical Poisson Point Process Intensity Function Modeling and Estimation with Measure Transport} 

\author{Tin Lok James Ng$^1$}
\author{Andrew Zammit-Mangion$^2$ \\ \vspace{0.2in} $^1$School of Computer Science and Statistics, Trinity College Dublin, Ireland \\$^2$School of Mathematics and Applied Statistics, University of Wollongong, Australia}





\begin{abstract}
  Recent years have seen an increased interest in the application of methods and techniques commonly associated with machine learning and artificial intelligence to spatial statistics. Here, in a celebration of the ten-year anniversary of the journal \emph{Spatial Statistics}, we bring together normalizing flows,  commonly used for density function estimation in machine learning, and spherical point processes, a topic of particular interest to the journal's readership, to present a new approach for modeling non-homogeneous Poisson process intensity functions on the sphere. The central idea of this framework is to build, and estimate, a flexible bijective map that transforms the underlying intensity function of interest on the sphere into a simpler, reference, intensity function, also on the sphere. Map estimation can be done efficiently using automatic differentiation and stochastic gradient descent, and uncertainty quantification can be done straightforwardly via nonparametric bootstrap. We investigate the viability of the proposed method in a simulation study, and illustrate its use in a proof-of-concept study where we model the intensity of cyclone events in the North Pacific Ocean. Our experiments reveal that normalizing flows present a flexible and straightforward way to model intensity functions on spheres, but that their potential to yield a good fit depends on the architecture of the bijective map, which can be difficult to establish in practice.
\end{abstract}




\end{frontmatter}


\section{Introduction}\label{sec:introduction}

The journal \emph{Spatial Statistics} has attracted contributions in a large range of topics  related to the discipline since its inception ten years ago. One of the topics that has featured consistently during its lifetime is that on point processes: between 10\% and 30\% of the articles published in the journal every year contained the word `point[-]process' or `point pattern' in either the title, the abstract, or the keywords. On the other hand, other topics have only recently began to feature in submissions, notably those related to machine learning and artificial intelligence (ML \& AI). For example, six articles contained the words `neural network' or `random forest' in the year 2021 (7\% of all publications in 2021), more than in all the other years combined.

This trend is not unique to the journal \emph{Spatial Statistics}, and reflects an increased adoption and awareness of modeling and inferential strategies that have been developed in the ML \& AI community. While the predictive ability and the use of technological innovation of models and algorithms in ML \& AI is undisputed, statisticians have a lot to contribute to this field, particularly in the area of uncertainty quantification. Largely for historical reasons, statisticians also tend to be involved in applications that are different from those in the ML \& AI community, and are drivers of integrating these new techniques with more classical statistical approaches in areas such as official statistics, geophysics, and ecology \citep[e.g.,][]{McDermott_2019a, Schafer_2020}.

Articles that incorporate ML \& AI in spatial statistics include those of \citet{Gerber_2021} and \citet{Lenzi_2021}, who use neural networks for estimating parameters governing spatial processes; \citet{McDermott_2019a}, who use echo state networks for the modeling and forecasting of spatio-temporal phenomena; \citet{Zammit_2020}, who use convolution neural networks to model the dynamics of geophysical phenomena in a dynamic spatio-temporal model; and \citet{Siden_2020, Li_2020, Zammit_2021, Murakami_2021}, who use ideas and technologies in deep learning to model arbitrarily complex spatial covariances or data models in spatial applications. This list is by no means exhaustive, but further highlights the considerable recent awareness of the role ML \& AI has in solving some of the pertinent challenges in spatial statistics in the last few years.

In this article we home in on a topic that today is commonly associated with ML \& AI: normalizing flows, which find their origin in an optimization problem known as `optimal transport'. In the simplest application of a normalizing flow, one expresses a \emph{target density} (i.e., a density function we wish to model) $f_0(\cdot)$, via the change of variables formula,
\begin{equation}\label{eq:cov_formula}
f_0(\xvec) = f_1(T(\xvec))|\det(\nabla T(\xvec))|, \quad \xvec \in \mathcal{X},
\end{equation}
where $f_1(\cdot)$ is a \emph{reference density}, commonly a normal density function, $\nabla$ is the gradient operator, $T(\cdot)$ is an arbitrarily complex, bijective, differentiable, map, and $\mathcal{X}$ is  a subset of the Euclidean plane. 
In ML \& AI, normalizing flows are conventionally used in several applications involving density-function estimation \citep[e.g.,][]{Papamakarios_2021} and variational Bayes \citep[e.g.,][]{Rezende_2015}, however notions related to these flows, and optimal transport more generally, are also starting to make an appearance in spatial statistics. For example, \citet{Katzfuss_2021} use transport maps to model the joint distribution of a spatial process at a large number of locations; \citet{Ng2020} use these maps on spatial coordinates to model non-homogeneous spatial Poisson point processes; and \citet{Chen_2020} use a type of normalizing flow, coined the continuous-time normalizing flow, for modeling spatio-temporal point patterns from earthquake and epidemiological data. Often, uncertainty in the map is not quantified or assessed, but there are exceptions to this: \cite{Katzfuss_2021} model transport maps using Gaussian processes, while \citet{Ng2020} present a nonparametric bootstrap approach to generate prediction intervals of the estimated intensity function at any spatial location.

In addition to point processes and ML \& AI, at least one article every year since 2014 in \emph{Spatial Statistics} was concerned with modeling on the sphere. Methods for modeling on the sphere are important for modeling processes at a global scale, however work in this area has attracted less attention than the Euclidean case. Research on the modeling of point processes on the sphere is even scarcer, and much of the literature and applications concerning point processes is on the plane  \citep[e.g.,][]{Dias2008, Adams2009, Miranda2011, Illian2012, Zammit2012}, although there has been a marked increased interest in the spherical case in recent years \citep{Robeson2014, Lawrence2016, Moller2018}. 

In recognition of the substantial and diverse contributions \emph{Spatial Statistics} has made to the community in the past decade, in this work we propose a methodology that ties together several of the threads of research discussed above: an approach for modeling the intensity function of a non-homogeneous Poisson point process on the sphere using normalizing flows. The paper is structured as follows: Section 2 outlines background material related to measure transport on the plane and on the sphere; Section 3 exploits the notion of process densities in Poisson processes to develop a measure transport approach to model intensity functions on spheres; Section 4 provides illustrations of the method on both simulated and application data; and Section 5 concludes.

\section{Background}\label{sec:background}
In this section we give some background on measure transport for density-function estimation, which is required for the intensity-function estimation procedure outlined in Section~\ref{sec:modeling}. We first focus on the case of density-function estimation on the Euclidean space in Section \ref{sec:background_euclidean} before moving to the spherical case in Section \ref{sec:background_sphere}.

\subsection{Transport of Probability Measure on Euclidean Spaces}\label{sec:background_euclidean}
Given two probability measures $\mu_0(\cdot)$ and $\mu_1(\cdot)$ defined on spaces ${\cal X}$ and ${\cal Z}$, respectively, a transport map $T: {\cal X} \rightarrow {\cal Z}$ is said to push forward $\mu_0(\cdot)$ to $\mu_1(\cdot)$ if, for any Borel subset $B \subset {\cal Z}$,
\begin{eqnarray}
\label{pushforward_measure}
 \mu_1(B) = \mu_0(T^{-1}(B)) ,
\end{eqnarray}
where the inverse $T^{-1}(\cdot)$ is set valued; specifically, $ T^{-1}(\zvec) = \{ \xvec \in {\cal X}: T(\xvec) = \zvec \} .$ For an injective transport map $T(\cdot)$, \eqref{pushforward_measure} can be re-formulated as
\begin{eqnarray}
\label{pushforward_measure2}
\mu_1(T(A)) = \mu_0(A),
\end{eqnarray}
for any Borel subset $A \subset {\cal X}$. Suppose that ${\cal X}, {\cal Z} \subset \mathbb{R}^{d}$, and that the measures $\mu_0(\cdot), \mu_1(\cdot)$ are absolutely continuous with respect to the Lebesgue measure on $\mathbb{R}^{d}$, with densities $\intd \mu_0(\xvec) / \intd \xvec= f_0(\xvec) $ and $\intd \mu_1(\zvec) / \intd \zvec = f_1(\zvec) $, respectively. If the map $T(\cdot)$ is bijective with a differentiable inverse $T^{-1}(\cdot)$, we obtain the familiar change-of-variables formula shown in \eqref{eq:cov_formula}, which expresses a complicated probability density $f_0(\cdot)$ in terms of a simpler density $f_1(\cdot)$ and a transport map $T(\cdot)$.
\\\\
A transport map allows for the simulation of random variables from complicated probability distributions. Suppose our interest is to draw a sample from the probability measure $\mu_0(\cdot)$. Given a simpler measure $\mu_1(\cdot)$ that is easy to simulate from and a transport map $T(\cdot)$ that satisfies \eqref{pushforward_measure}, one can first simulate a sample $\Zvec$ from $\mu_1(\cdot)$, and then apply the inverse transport map $T^{-1}(\cdot)$ to $\Zvec$; the resulting sample $T^{-1}(\Zvec)$ will then be distributed according to $\mu_0(\cdot)$. The inverse transport map $T^{-1}(\cdot)$ may not be analytically available in general and, in such cases, numerical procedures for evaluating the inverse map are required.
\\\\
The measure transport approach presents an attractive way to estimate probability density functions as it only involves specifying a reference density $f_1(\cdot)$, which is typically chosen to be the density of the standard multivariate normal distribution, and a transport map $T(\cdot)$. \cite{Marzouk2016} discusses various strategies for parameterizing the transport map $T(\cdot)$, and more recent approaches typically parameterize $T(\cdot)$ using a neural network \citep[e.g.,][]{Papamakarios2017, Huang2018, Kobyzev_2020}. The bijectivity and differentiability properties of the map $T(\cdot)$ are preserved under compositions. That is, given two bijective maps $T^{(1)}(\cdot), T^{(2)}(\cdot)$ with differentiable inverses, their composition $T^{(2)} \circ T^{(1)}(\cdot)$ remains bijective with a differentiable inverse, and hence there is no issue of ``space folding.'' The Jacobian determinant of the resulting composition remains computationally tractable since, by the chain rule,
$$ \mbox{det}(\nabla (T^{(2)} \circ T^{(1)}(\xvec))) = \mbox{det}(\nabla T^{(1)}(\xvec)) \mbox{det}(\nabla T^{(2)} (T^{(1)}(\xvec)) ), \quad \xvec \in \mathcal{X}.$$
These properties associated with taking compositions facilitate the construction of complex transformations via the composition of multiple simpler transformations. Thus, an approach often used is to define $T(\cdot)$ as a composite of multiple transformations, that is, as $T(\cdot) \equiv T^{(K)} \circ \cdots \circ T^{(1)}(\cdot)$, where $T^{(k)}(\cdot)$ transforms $\zvec^{(k-1)}$ into $\zvec^{(k)}$, with $\zvec^{(0)} \equiv \xvec$ and $\zvec^{(K)} \equiv \zvec$. This `flow' of transformations leads to the class of maps known as normalizing flows which, as discussed in Section~\ref{sec:introduction}, are widely used in ML \& AI. 
Here, the problem of density-function estimation is equivalent to that of estimating the unknown parameters of the transport maps $T^{(1)}(\cdot),\dots,T^{(K)}(\cdot)$, which can be efficiently performed using automatic differentiation libraries, stochastic gradient descent and, with maps involving neural networks, graphics processing units.

 \subsection{Transport of Probability Measure on Spheres} \label{sec:background_sphere}
While most of the literature on measure transport and normalizing flows to date has concentrated on Euclidean spaces, some recent work has focused on normalizing-flow techniques for density-function estimation on Riemannian manifolds such as spheres and tori  \citep{Gemici2016, Mathieu2020, Rezende2020}. \cite{Gemici2016} developed a general approach for probability density-function estimation for Riemannian manifolds by first projecting the manifold to the Euclidean space $\mathbb{R}^{d}$, applying conventional Euclidean normalizing flows in this space, and then  projecting $\mathbb{R}^{d}$ back to a manifold that is not necessarily the same as the original manifold. However, if the manifolds are not diffeomorphic to $\mathbb{R}^{d}$, as in the case of spheres, singularities arise that can lead to numerical instabilities during parameter estimation. To circumvent this difficulty, normalizing flows tailored for Riemannian manifolds have been developed; these include exponential map flows, recursive M{\"o}bius spline flows \citep{Rezende2020}, and Riemannian continuous normalizing flows \citep{Mathieu2020}.
\\\\
In this paper we focus on the exponential map flow proposed by \citet{Rezende2020}, which is based on a mechanism for establishing density functions on the sphere first proposed by \citet{Sei2013}, which in turn builds on the theoretical results of \citet{McCann2001}. The core idea is to construct a transport map from the space of \emph{wrapping potential functions}, $W(\mathbb{S}^{d-1})$, $d \ge 2$, which contains differentiable functions on the sphere that exhibit a property known as $c$-convexity \citep[see Definition 3.1 in ][for a formal definition]{Sei2013}.  Consider, for now, a single wrapping potential function $\phi(\cdot) \in W(\mathbb{S}^{d-1})$. A valid transport map $G_{\phi}: \mathbb{S}^{d-1} \rightarrow \mathbb{S}^{d-1}$ can be constructed by taking the \emph{exponential map} of the gradient $\nabla \phi(\cdot) \in T_{\xvec} \mathbb{S}^{d-1}$ where, with a slight abuse of notation, $T_{\xvec} \mathbb{S}^{d-1}$ is the tangent space of $\mathbb{S}^{d-1}$ at $\xvec$. (An exponential map $\exp_{\xvec}(\vvec), \vvec \in T_{\xvec}\mathbb{S}^{d-1}$, not to be confused with the exponential operator, returns the end-position of a particle that starts off at $\xvec$ and travels for one time unit with velocity $\vvec$ on $\mathbb{S}^{d-1}$.) Specifically, given a wrapping potential function $\phi(\cdot)$,
\begin{eqnarray}
\label{exp_map_flow}
 T(\xvec) = G_{\phi}(\xvec) = \exp_{\xvec}(\nabla \phi(\xvec)), \quad \xvec \in \mathbb{S}^{d-1}, 
\end{eqnarray}
is a valid transport map, in the sense that it is homeomorphic and locally diffeomorphic. Interestingly, while the exponential map $G_\phi(\cdot)$ is generally analytically intractable for arbitrary Riemannian manifolds, a closed form expression for  $\exp_{\xvec}: T_{\xvec} \mathbb{S}^{d-1} \rightarrow \mathbb{S}^{d-1}$ in Equation \eqref{exp_map_flow} is available:
\begin{equation}\label{eq:expmap}
  \exp_{\xvec}(\vvec) = (\cos \| \vvec \|) \xvec + (\sin \|\vvec\| )(\vvec/\|\vvec\|) ,\quad \vvec \in T_{\xvec} \mathbb{S}^{d-1},
  \end{equation}
where $\|\cdot\|$ is the Euclidean norm. 
One can therefore readily construct a probability distribution on the sphere with density function
$$ f_{\phi}(\xvec) \propto |\det(\nabla G_{\phi}(\xvec))|, \quad \xvec \in \mathbb{S}^{d-1},$$
where we have taken the uniform measure as the reference measure, and where the Jacobian is taken with respect to any orthonormal basis on $T_{\xvec} \mathbb{S}^{d-1}$ and $T_{G_{\phi}(\xvec)} \mathbb{S}^{d-1}$. Simulating random variables from this probability distribution is straightforward and involves first sampling $\Zvec \sim \mbox{Unif}(\mathbb{S}^{d-1})$, the uniform distribution on $\mathbb{S}^{d-1}$, and then solving $\Xvec = G_{\phi}^{-1}(\Zvec)$. Efficient methods exist for drawing samples from $\mbox{Unif}(\mathbb{S}^{d-1})$; for example, one could first sample from the standard $d$-dimensional Gaussian distribution, and then normalize the sample so that each $d$-dimensional vector has a magnitude of one. The equality $ \Xvec = G_{\phi}^{-1}(\Zvec)$ can also be solved efficiently \citep{McCann2001}. The exponential map flow offers a simple, intuitive, and flexible approach to modeling density functions on $\mathbb{S}^{d-1}$, since one only needs to specify a wrapping potential function $\phi(\cdot)$ for its construction. 
\\\\
The space of wrapping potential functions $W(\mathbb{S}^{d-1})$ has the desirable property of closure under convex combinations. That is, if $\phi_i(\cdot) \in W(\mathbb{S}^{d-1})$, for $i=1,\ldots,p$, then $\sum_{i=1}^{p} \eta_i \phi_i(\cdot) \in W(\mathbb{S}^{d-1})$ if $\eta_i > 0$ and $\sum_{i=1}^{p} \eta_i = 1$. This property facilitates the construction of flexible families of probability distributions via relatively simple functions $\{\phi_i(\cdot)\}$. One general approach to construct $\phi(\cdot) \in W(\mathbb{S}^{d-1})$ is described in \cite{Sei2013}: Let $\mvec_1, \ldots, \mvec_p \in \mathbb{S}^{d-1}$, and $\phi_1(\cdot), \ldots, \phi_p(\cdot) \in W(\mathbb{S}^{d-1})$, and define
\begin{eqnarray}
\label{wrapping_potential_func}
 \phi(\xvec) \equiv \sum_{i=1}^{p} \eta_i \phi_i(d(\xvec, \mvec_i)), \quad \xvec \in \mathbb{S}^{d-1},
\end{eqnarray}
where $d(\xvec,\yvec)$ is the geodesic distance between $\xvec,\yvec \in \mathbb{S}^{d-1}$, and $\eta_i > 0$, $\sum_{i=1}^p \eta_i = 1$. In some cases, a closed form expression for the resulting probability density function $f_{\phi}(\cdot)$ is then available, which facilitates practical implementation. Various tractable specifications of the function $\phi_i(\cdot)$ are proposed in \cite{Sei2013} and \citet{Rezende2020}. In this paper we adopt the radial flow proposed in \cite{Rezende2020} where $ \phi_i(\xvec) =  e^{\beta_i (\cos d(\xvec,\mvec_i) - 1)}/\beta_i ,$ and, hence, 
\begin{eqnarray}
\label{radial_flow}
 \phi(\xvec) = \sum_{i=1}^{p} \frac{\eta_i}{\beta_i} e^{\beta_i (\cos d(\xvec, \mvec_i) - 1 ) }, \quad \xvec \in \mathbb{S}^{d-1},
\end{eqnarray}
where, for $i = 1,\dots,p$,  $\beta_i >0, \mvec_i \in \mathbb{S}^{d-1}$, and $\eta_i > 0$ ($\sum_{i=1}^{p} \eta_i = 1$) are model parameters that need to be estimated.

\begin{figure}[t!]
  \includegraphics[width=\linewidth]{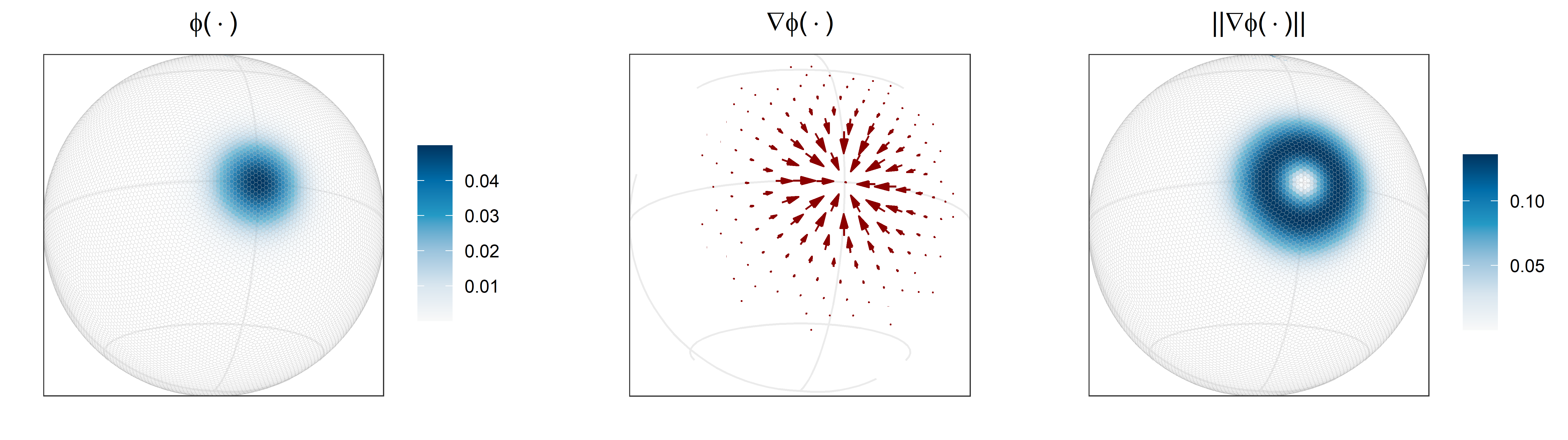}
  \caption{Left panel: The wrapping potential $\phi(\cdot)$ evaluated over an icosahedral Snyder equal area aperture 3 hexagon (ISEA3H) discrete global grid \citep[DGG,][]{Sahr_2003} at resolution 7. Middle panel: The gradient field $\nabla\phi(\cdot)$ evaluated at the centroids of an ISEA3H DGG at resolution 4 with arrow sizes proportional to the vector norms. Right panel: The gradient norm $\|\nabla\phi(\cdot)\|$ evaluated over an ISEA3H DGG at resolution 7.\label{fig:phi_plots}}
\end{figure}

\begin{figure}[t!]
  \includegraphics[width=\linewidth]{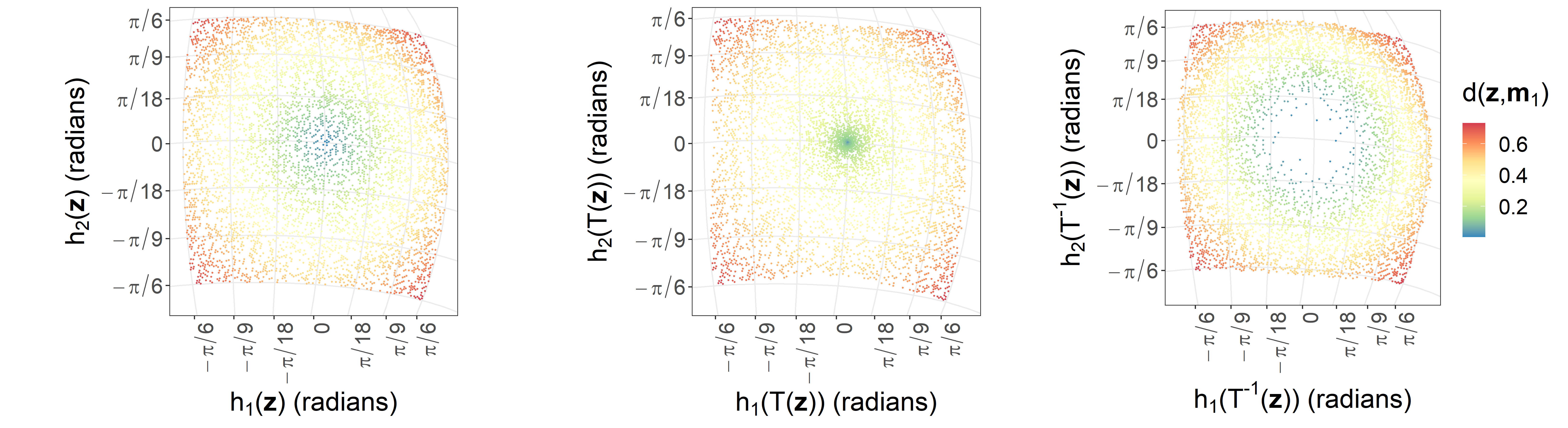}
  \caption{Left panel: A sample $\Zvec$ of 5000 points from a uniform distribution on the surface of the sphere in the region $[-\pi/6, \pi/6] \times [-\pi/6, \pi/6]$. Middle panel: The application of the exponential map $T(\cdot)$ to $\Zvec$. Right panel: The application of the inverse exponential map, $T^{-1}(\cdot)$, to $\Zvec$. In all panels, the colour is used to represent the geodesic distance of a point in the sample $\Zvec$ from $\mvec_1 = (1, 0,0)'$, with blue, yellow, and red representing increasing distances, respectively, and the function $h(\cdot)$ maps from Cartesian to spherical coordinates. \label{fig:points_plots}}
\end{figure}

To help visualize the action of measure transport on spheres we give a small illustration. Consider \eqref{radial_flow} with $p = \eta_1 = 1$, $\beta_1 = 20$, and $\mvec_1 = (1,0,0)'$. This function, depicted in the left panel of Figure~\ref{fig:phi_plots}, has a peak at $\mvec_1$ and decays radially from this point. The vector field $\nabla\phi(\cdot)$ and its norm are shown in the middle and right panels of Figure~\ref{fig:phi_plots}, respectively. To illustrate the behavior of the exponential map of $\nabla\phi(\cdot)$, we simulate a random sample $\Zvec$ comprising 5000 points on the angular sub-domain $[-\pi/6,\pi/6] \times [-\pi/6,\pi/6]$ radians, and compute the terminal destination of these points after a time unit on spherical trajectories dictated by $\nabla\phi(\cdot)$, by evaluating \eqref{eq:expmap} for each of these points. The uniformly-distributed points, and the exponential map of these points, are shown in the left and middle panels of Figure~\ref{fig:points_plots}, respectively. Note how the exponential map has the effect of pulling elements in $\Zvec$ toward $\mvec_1$. Now, recall that the exponential map is used to relate our target variable $\Xvec$ to the reference variable $\Zvec$ via the mapping $T(\Xvec) = \Zvec$, and that therefore a random sample of $\Xvec$ is obtained by applying $T^{-1}(\cdot)$ to $\Zvec$. Intuitively, this inverse finds the initial departure location of points for which we know the terminal location. The effect is thus opposite to what we observe when applying $T(\cdot)$ to $\Zvec$, and the inverse map causes elements in $\Zvec$ to diverge away from $\mvec_1$, as seen in the right panel of Figure~\ref{fig:points_plots}.

We conclude this section by noting that, as in the Euclidean case, the flexibility of the resulting probability density function can be further improved by taking compositions of multiple transport maps: 
\begin{eqnarray}
\label{mt_comp}
 G_{\phi}^{(K)} \circ \cdots G_{\phi}^{(1)}(\xvec),\quad \xvec \in \mathbb{S}^{d-1}, 
\end{eqnarray}
where each $G_\phi^{(k)}(\cdot), k=1, \ldots,K$, has the form given in \eqref{exp_map_flow}.

\section{Intensity Function Modeling with Measure Transport}\label{sec:modeling}
In this section we apply the theory outlined in Sections~\ref{sec:background_euclidean} and \ref{sec:background_sphere} to intensity function estimation of a non-homogeneous point process (NHPP).
\\\\
A point process $X$ on a domain ${\cal X}$ is a random finite subset of ${\cal X}$ with corresponding counting measure $N(\cdot)$, where for any subset $A \subset \cal X$, $N(A)$ denotes the number of points in $X$ falling in $A$. For a homogeneous Poisson process with rate $\lambda$, $N(A)$ is a Poisson random variable with mean $\lambda|A|$, where $|A|$ is the area of the set $A$, and the random variables $N(A_1), N(A_2)$ are independent for disjoint regions $A_1, A_2$. An NHPP defined on ${\cal X}$ is a Poisson point process with varying intensity, which is fully characterized by its intensity function $\lambda: {\cal X} \rightarrow [0,\infty)$. NHPPs have mostly (but not exclusively) been modeled and fitted on Euclidean domains, that is, where ${\cal X} \subset \mathbb{R}^{d}$ and $d \ge 1$.
\\\\
The inferential target when modeling NHPPs is the underlying intensity function. Nonparametric techniques, such as spline-based methods \citep{Dias2008} and wavelet-based methods \citep{Miranda2011},  which do not fix the functional form of the intensity function, are among the many popular approaches for intensity function estimation. However, these methods generally do not scale well with the number of observed points or the dimension $d$. Model-based approaches, such as those using a doubly-stochastic Poisson process \citep{Moller1998} are also popular. However, inference with these models is typically computationally intensive, and often one needs to resort to approximate inferential methods for computations to remain tractable. 
Recently, \cite{Ng2020} proposed a measure transport approach to model the intensity function of an NHPP on the Euclidean domain based on normalizing flows. The proposed approach was shown to be competitive with kernel-based methods and log-Gaussian Cox process models on simulation studies and a data application. In Section \ref{sec:mt_euclidean} we review this measure transport approach for intensity function estimation on $\mathbb{R}^d$. Section \ref{sec:mt_sphere} then extends the method to the spherical domain. 

\subsection{Intensity Function Modeling on $\mathbb{R}^{d}$ with Measure Transport}\label{sec:mt_euclidean}
Consider a NHPP defined on ${\cal X} \subset \mathbb{R}^{d}$ with intensity function $\lambda(\cdot)$. The log-likelihood function of $\lambda(\cdot)$ with respect to a unit rate Poisson process can be written as \cite[e.g.,][]{Moller1998}, 
\begin{eqnarray}
\label{obj_func1}
 \ell(\lambda; \{\xvec_i\}_{i=1}^n) = - \int_{{\cal X}} (\lambda(\xvec) - 1 ) \intd \xvec + \sum_{i=1}^{n} \log \lambda(\xvec_i),
\end{eqnarray}
where $\{\xvec_i\}_{i=1}^{n}$ is a realization of the point process, and $n$ is the number of points in the realization. Instead of optimizing the likelihood function in \eqref{obj_func1} directly, the core idea introduced by \cite{Ng2020} is to model the so called ``process density,'' $\tilde{\lambda}(\cdot) = \lambda(\cdot) / \mu_{\lambda}({\cal X}) $  using a measure transport approach, where $\mu_\lambda(\cal X)$ is the integrated intensity defined as
$$ \mu_{\lambda}({\cal X}) = \int_{ {\cal X}} \lambda( \xvec ) \intd \xvec   .$$
The quantity $\tilde{\lambda}(\cdot)$ can be regarded as a density function associated with the point process with respect to the Lebesgue measure on ${\cal X}$. The process density representation was introduced by \cite{Taddy2010}; this representation is attractive as it allows the process density $\tilde{\lambda}(\cdot)$ to be modeled separately from $\mu_{\lambda}({\cal X})$. 
\\\\
\citet{Ng2020} model the process density $\tilde{\lambda}(\cdot)$ using the transport map $T: {\cal X} \rightarrow {\cal Z}$, where ${\cal Z}$ need not be the same as ${\cal X}$. The map $T(\cdot)$ is then constructed from a class of normalizing flows known as neural autoregressive flows, which can approximate a large class of process densities arbitrarily well \citep{Huang2018, Ng2020}. Specifically, it can be shown that the class of neural autoregressive flows has the  desirable ``universal approximation property.'' Choosing a simple reference density function $\xi(\cdot)$ on ${\cal Z}$ such as the standard multivariate normal distribution if ${\cal Z} = \mathbb{R}^{d}$, and the uniform density if ${\cal Z}$ is bounded, the process density $\tilde{\lambda}(\cdot)$ can be expressed as 
$$\tilde{\lambda}(\xvec; \Theta) = \xi(T(\xvec; \Theta)) |\det (\nabla T(\xvec; \Theta))|, \quad \xvec \in {\cal X} .$$ 
where we have now explicitly denoted the dependence of $T(\cdot)$, and hence that of $\tilde{\lambda}(\cdot)$, on parameters $\Theta$ used in constructing the normalizing flow and that need to be estimated.  The log-likelihood function of the parameters is  simply the log of the process density, and is given by \begin{eqnarray}
\label{lik_func}
  \ell(\Theta; \{\xvec_i\}_{i=1}^n)  =  \sum_{i=1}^{n} \bigg( \log \xi(T(\xvec_i; \Theta)) + \log |\det (\nabla T(\xvec_i; \Theta)) | \bigg) .
\end{eqnarray}
Maximum likelihood estimation of $\Theta$, and hence of $\tilde{\lambda}(\cdot)$, reduces to numerical optimization of the log-likelihood function \eqref{lik_func} with respect to $\Theta$.   Further, since $N(\cal X)$ is a Poisson random variable with mean $\int_{\cal X} \lambda(\xvec) \intd \xvec = \mu_{\lambda}(\cal X)$, $ \mu_{\lambda}(\cal X)$ can be estimated by $n$, the number of points of the realized point process. Therefore, an estimate for the intensity function $\lambda(\cdot)$ is given by $ \hat{\lambda}(\cdot) = n \hat{ \tilde{\lambda}}(\cdot)$, where $ \hat{ \tilde{\lambda}}(\cdot) \equiv \tilde\lambda(\cdot\,; \hat\Theta)$ is the estimated process density, and $\hat\Theta$ is the maximum likelihood estimate of $\Theta$.

\subsection{Intensity Function Modeling on $\mathbb{S}^{d-1}$ with Measure Transport}
\label{sec:mt_sphere}

In this section we consider the problem of estimating the intensity function of a NHPP on $\mathbb{S}^{d-1} \equiv \{ \xvec \in \mathbb{R}^{d}: \|\xvec\| = 1\}$ where $\|\cdot\|$ is the $L^2$ norm, for $d \ge 2$ but with a particular focus on $\mathbb{S}^{2}$.  An NHPP on $\mathbb{S}^{d-1}$ is characterized by its intensity function $\lambda(\xvec), \xvec \in \mathbb{S}^{d-1}$, where $N(A)$ is a Poisson random variable with mean $\int_{A} \lambda(\xvec) \intd \xvec $ for $A \subset \mathbb{S}^{d-1}$. We extend the measure transport approach to intensity function modeling on the Euclidean space described in Section \ref{sec:mt_euclidean} to the spherical domain by employing the normalizing flows described in Section \ref{sec:background_sphere}. 
\\\\
Our objective is to estimate the intensity function $\lambda(\cdot)$ based on the realized points $\{\xvec_i\}_{i=1}^n$. Analogously to the Euclidean case, the log-likelihood function of $\lambda(\cdot)$ with respect to the unit rate Poisson process on the sphere can be written as
\begin{eqnarray}
\label{obj_func2}
 {\cal \ell}(\lambda; \{\xvec_i\}_{i=1}^n) = - \int_{\mathbb{S}^{d-1}} (\lambda(\xvec) - 1 ) \intd \xvec + \sum_{i=1}^{n} \log \lambda(\xvec_i) .
\end{eqnarray}
As in Section \ref{sec:mt_euclidean}, we define the process density as $\tilde{\lambda}(\cdot) = \lambda(\cdot) / \mu_{\lambda}(\mathbb{S}^{d-1}) $, where $\mu_{\lambda}(\mathbb{S}^{d-1}) = \int_{\mathbb{S}^{d-1}} \lambda(\xvec) \intd \xvec $ is the integrated intensity, and separately estimate $\mu_{\lambda}(\mathbb{S}^{d-1})$ and $\tilde{\lambda}(\cdot)$. Since $N(\mathbb{S}^{d-1})$ is a Poisson random variable with mean $\int_{\mathbb{S}^{d-1}} \lambda(\xvec) \intd \xvec = \mu_{\lambda}(\mathbb{S}^{d-1})$, $ \mu_{\lambda}(\mathbb{S}^{d-1})$ can be estimated by $n$, the number of points in the realization.  Since $\int_{\mathbb{S}^{d-1}} \tilde{\lambda}(\xvec) \intd \xvec = 1$, $\tilde{\lambda}(\cdot)$ is a valid probability density on $\mathbb{S}^{d-1}$ with respect to the Lebesgue measure, and the measure transport framework described in Section \ref{sec:background_sphere} can be applied. 
\\\\
We apply the exponential map radial flow given by \eqref{exp_map_flow} and \eqref{radial_flow} to model the process density $\tilde{\lambda}(\cdot)$. We use the uniform measure as the reference probability measure on $\mathbb{S}^{d-1}$; then, the log-likelihood function \eqref{lik_func} simplifies to
\begin{eqnarray}
\label{obj_func3}
    \ell(\Theta; \{\xvec_i\}_{i=1}^n)  = \sum_{i=1}^{n} \log |\det(\nabla G_{\phi}(\xvec_i; \Theta))| + \textrm{const.},
\end{eqnarray}
where we have now explicitly denoted the dependence of $G_{\phi}(\cdot)$ on the model parameters $\Theta \equiv  \{ \beta_i^{(k)}, \mvec_i^{(k)}, \eta_i^{(k)}: k=1, \ldots, K; i=1,\ldots,p \} $ that construct a composition of $K$ radial flows, where $G_{\phi}(\cdot) \equiv G_{\phi}^{(K)} \circ \cdots \circ G_{\phi}^{(1)}(\cdot)$ is the composition of $K$ exponential maps with radial flow, and where the $k$-th map in the composition has model parameters $\{ \beta_i^{(k)}, \mvec_i^{(k)}, \eta_i^{(k)} \}_{i=1}^{p}$. The simplification arises from our choice of the uniform measure as the reference measure, which leads to the first term in the summation of the log-likelihood function \eqref{lik_func} to be constant (denoted as `const.' in \eqref{obj_func3}). Estimating the process density $\tilde{\lambda}(\cdot)$ proceeds by (numerically) optimizing the log-likelihood function in \eqref{obj_func3} with respect to $\Theta$. 
\\\\
To complete the specification of the model, we need to choose the number of maps in the composition, $K$, and the number of basis functions constructing each radial flow in \eqref{radial_flow}, $p$. We found that a small value of $p$ (e.g. $p=1$ or $p=2$) and a moderate to large value of $K$ (between 20 and 40) led to good performance in our simulation experiments. This observation is consistent with much of the literature in deep learning \citep[e.g.,][]{Eldan2016, Raghu2017}, where deep, narrow, neural networks are seen to have more representational efficiency than shallow, wide, networks. A theoretic rationale for choosing the optimal values of $K$ and $p$ is unfortunately not available. Their selection is analogous to the problem of choosing the optimal number of hidden layers and nodes per layer for deep neural network models, which is often solved by training a large number of models on training data, and testing them on (unseen) test data. We optimize \eqref{obj_func3} by  using automatic differentiation libraries and stochastic gradient descent in \emph{PyTorch} \citep{Paszke2017}. The estimated process density $\hat{\tilde{\lambda}}(\cdot)$ is then obtained from the estimated model parameters, and finally the estimated intensity function $\hat{\lambda}(\cdot)$ is simply given by $n \hat{\tilde{\lambda}}(\cdot)$.
\\\\
To quantify the uncertainty on the estimated intensity function we proceed as in \cite{Ng2020} and adopt a nonparametric bootstrapping approach \citep{Efron1981}. This approach requires the sampling of $B$ bootstrap samples, which are generated as follows: First, the number of points $n_b, b=1,\ldots, B$, is simulated from a Poisson distribution with rate parameter $n$. Second, for each bootstrap sample $b=1,\ldots,B$, $n_b$ points from the observed points $\{\xvec_i\}_{i=1}^n$ are sampled with replacement. Third, $B$ process densities, $\hat{\tilde{\lambda}}_b(\cdot), b=1, \ldots, B,$ are estimated using the above approach, yielding the estimated intensity functions $\hat{\lambda}_b(\cdot) = n_b \hat{\tilde{\lambda}}_b(\cdot), b = 1,\dots,B$. Finally, we quantify the uncertainty at any point $\xvec \in \mathbb{S}^{d-1}$ through percentiles of $\{\hat{\lambda}_b(\xvec): b=1, \ldots, B\}$.

\section{Simulation Studies and Application}\label{sec:experiments}
In this section we assess the performance of our proposed method through simulation studies (Section~\ref{sec:sim}) and an application where the aim is to model the intensity of cyclone events in the North Pacific Ocean (Section~\ref{sec:cyclones}).

\subsection{Simulation Experiment}\label{sec:sim}
\begin{figure*}[!t]

\centering
 \includegraphics[width=\linewidth]{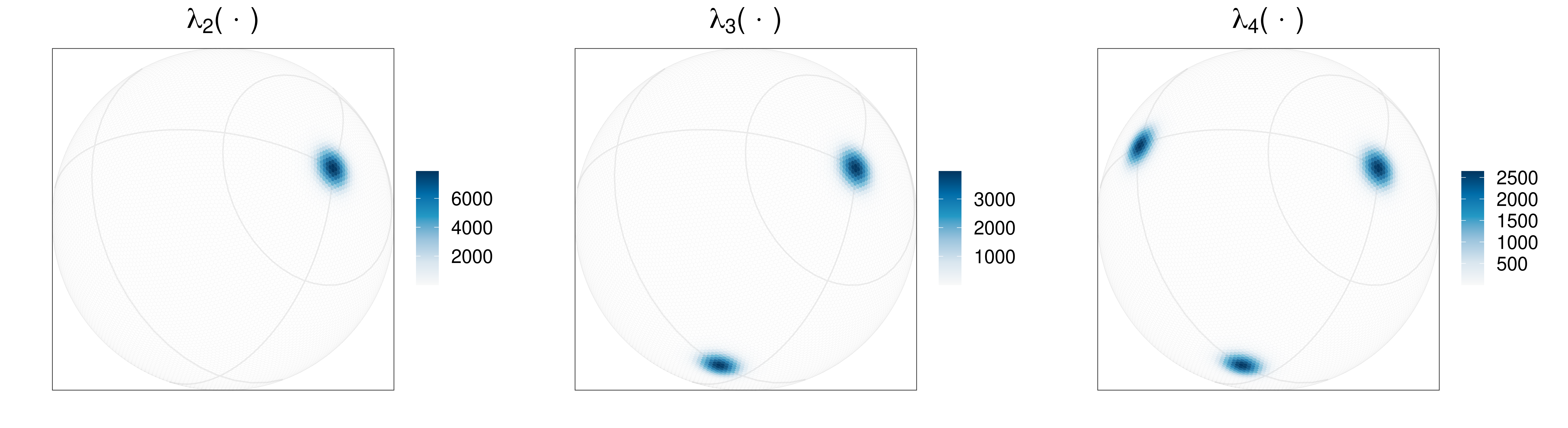}

  \caption{Intensity functions constructed via mixtures of von Mises--Fisher density functions that are used in the simulation experiment of Section~\ref{sec:sim}.  Left panel: The intensity function $\lambda_2(\cdot)$. Center panel: The intensity function $\lambda_3(\cdot)$. Right panel: The intensity function $\lambda_4(\cdot)$.}
\label{vMF_plot}
\end{figure*}

In this section we carry out an experiment to assess the accuracy of the measure transport approach of Section \ref{sec:mt_sphere} when estimating known intensity functions from point-process realizations, and also to assess the sensitivity of the estimates to the number of transport maps used in the composition of the normalizing flows.
\\\\
In our simulation experiment, Poisson point-process realizations were simulated from the following intensity functions:
\begingroup
\allowdisplaybreaks
\begin{eqnarray}
 \lambda_1(\xvec) &=& n f_{unif}(\xvec), \label{eq:lambda1}\\
 \lambda_2(\xvec) &=& n f_{vmf}(\xvec; \mvecc, \kappa), \\
 \lambda_3(\xvec) &=& \pi n f_{vmf}(\xvec; \mvecc_1, \kappa_1) + (1 - \pi) n f_{vmf}(\xvec; \mvecc_2, \kappa_2), \\
 \lambda_4(\xvec) &=& \pi_1 n f_{vmf}(\xvec; \mvecc_1, \kappa_1) + \pi_2 n f_{vmf}(\xvec; \mvecc_2, \kappa_2) + \nonumber \\ && (1-\pi_1-\pi_2) n f_{vmf}(\xvec; \mvecc_3, \kappa_3), \label{eq:lambda4}
\end{eqnarray}
\endgroup
for $\xvec \in \mathbb{S}^2$, where $f_{unif}(\cdot)$ denotes the uniform density on $\mathbb{S}^{2}$; $f_{vmf}(\cdot\,; \mvecc, \kappa)$ denotes the von Mises--Fisher distribution on $\mathbb{S}^{2}$ with location parameter $\mvecc$ and concentration parameter $\kappa$; $\pi, \pi_1, \pi_2 \in [0,1]$; and $\pi_1 + \pi_2 \in [0,1]$. We set $n=500$ for all intensity functions. For $\lambda_2(\cdot)$, we set $\mvecc = (0,0,1)'$ and $\kappa=100$. For $\lambda_3(\cdot)$, we set $\mvecc_1 = (0,1,0)', \mvecc_2 = (0,0,1)', \kappa_1 = \kappa_2 = 100,$ and $\pi = 0.5$. For $\lambda_4(\cdot)$, we set $\mvecc_1 = (1,0,0)', \mvecc_2 = (0,1,0)', \mvecc_3 = (0,0,1)'$, $\kappa_1 = \kappa_2 = \kappa_3 = 100$, and $\pi_1 = \pi_2 = 1/3$. We considered a range of values for the number of transport maps $K$, and we set $p=1$ for the radial flow in \eqref{radial_flow}. We chose $f_{unif}(\cdot)$ as the reference density on $\mathbb{S}^{2}$.  \cite{Rezende2020} showed, in the density estimation setting, that density functions constructed from a mixture of von Mises--Fisher density functions can be well estimated using exponential map radial flows. Therefore, the main purpose of our simulation experiment is to assess the sensitivity of the method to different values of $K$, and to compare the quality of fits to that obtained using a log-Gaussian Cox process (LGCP) model-based approach to intensity function estimation. The intensity functions $\lambda_2(\cdot), \lambda_3(\cdot)$ and $\lambda_4(\cdot)$ are visualized in Figure~\ref{vMF_plot}.\\\\

For each intensity function ($\lambda_1(\cdot),\dots,\lambda_4(\cdot)$), we generated 20 independent point-process realizations, and fit intensity functions to each of the realizations using transport maps with $K \in \{1,5,10,20,30,40 \}$.  We assessed goodness of fit through the $L^1$ distances between the true and the estimated intensity functions, for each of the four cases. We also compared the goodness-of-fit metrics obtained from our proposed method to those from an LGCP model with a Mat{\'e}rn covariance function with smoothness parameter equal to one. The LGCP was fitted to the data using the package \texttt{inlabru} \citep{Bachl2019} using an irregular finite-element tessellation of the sphere comprising approximately 2000 elements. This number of elements is sufficient given that $\lambda_1(\cdot),\dots,\lambda_4(\cdot)$ vary relatively slowly (and not at all in the case of $\lambda_1(\cdot)$) spatially. The average and empirical standard deviations of the $L^1$ distances across the 20 realizations were computed for each case; the results are shown in Table \ref{simall_table}.
\\\\
From Table \ref{simall_table}, we see that the proposed method considerably underperforms the LGCP when the true intensity function is uniform on $\mathbb{S}^2$ $(\lambda_1(\cdot))$, despite this also being the reference density for the normalizing flow. Further, the intensity-function estimate deteriorates as the number of compositions of transport maps, $K$, increases, and $K = 1$ (the least complex map) gives by far the best fit. This highlights a limitation of our approach with radial flow, which does not easily model the identity map due to its functional form; this lies in contrast with other warping approaches that allow the identity map to be easily inferred \citep[e.g.,][]{Zammit_2021}. On the other hand, when the true intensity functions are more variable ($\lambda_2(\cdot)$--$\lambda_4(\cdot)$), we find cases where transport maps begin to yield a better fit to the true underlying intensity function than a classic LGCP. In these examples, we find that $K = 20$--$30$ compositions (i.e., 20--30 radial flows) with $p = 1$ (i.e., one radial flow per compositional layer) give the best fits. These results show that, as is often the case with models constructed via composition such as deep learning models, the optimal number of layers to choose ($K$), and the width of each layer ($p$), is a non-trivial problem. This is especially the case here, where cross-validation (as is often done for model selection in this scenario) does not easily generalize to the point process case.

\begin{table}[t!]
\caption{Average and empirical standard deviation of the 20 $L^1$ distances between the true and the fitted intensity functions in the simulation experiment for each of $\lambda_1(\cdot),\dots,\lambda_4(\cdot)$ as given in \eqref{eq:lambda1}--\eqref{eq:lambda4}. Boldface denotes the best performance in an $L^1$ sense for each case.\label{simall_table}}
\footnotesize
\begin{center}
\begin{tabular}{c|lcccccc|c}
   \hline
   Case & No. of compositions ($K$)  & 1 & 5& 10 & 20 &30 &  40 & LGCP  \\ \hline\hline
  
\multirow{ 2}{*}{$\lambda_1(\cdot)$} & Average $L^1$ distance & 10.63 & 23.45 & 24.43 & 25.37 & 28.11 & 28.67  & \textbf{5.01} \\ 
   & Stdev. of $L^1$ distance & 2.07 & 2.87 &1.88 & 1.89 & 4.40 & 3.17 & 1.89 \\\hline
\multirow{ 2}{*}{$\lambda_2(\cdot)$} & Average $L^1$ distance & 68.83 & 52.56 & 21.52 & \textbf{6.62} & 16.11 & 22.41 & 11.61 \\ 
 & Stdev. of $L^1$ distance & 2.08 & 1.82 &2.39 & 1.48 & 7.12 & 6.03 & 1.78 \\ \hline   
\multirow{ 2}{*}{$\lambda_3(\cdot)$} & Average $L^1$ distance & 69.70& 56.45 & 34.81 & \textbf{11.24} & 14.51 & 15.67 & 13.90 \\ 
& Stdev. of $L^1$ distance & 1.00 & 3.55 & 4.92 & 2.56 & 2.49 & 2.21 & 1.78 \\ \hline 
\multirow{ 2}{*}{$\lambda_4(\cdot)$} & Average $L^1$ distance & 65.93 & 52.01 & 40.96 & 22.82 & \textbf{12.35} & 16.45 & 14.71 \\ 
& Stdev. of $L^1$ distance & 1.05 & 2.79 & 2.52 & 3.60 & 2.06 & 2.35 & 1.49 \\    \hline\hline 
\end{tabular}
\end{center}
\end{table}

\subsection{Proof of Concept: Application to Cyclone Data}\label{sec:cyclones}
\begin{figure*}[!t]

\centering

 \includegraphics[width=\linewidth]{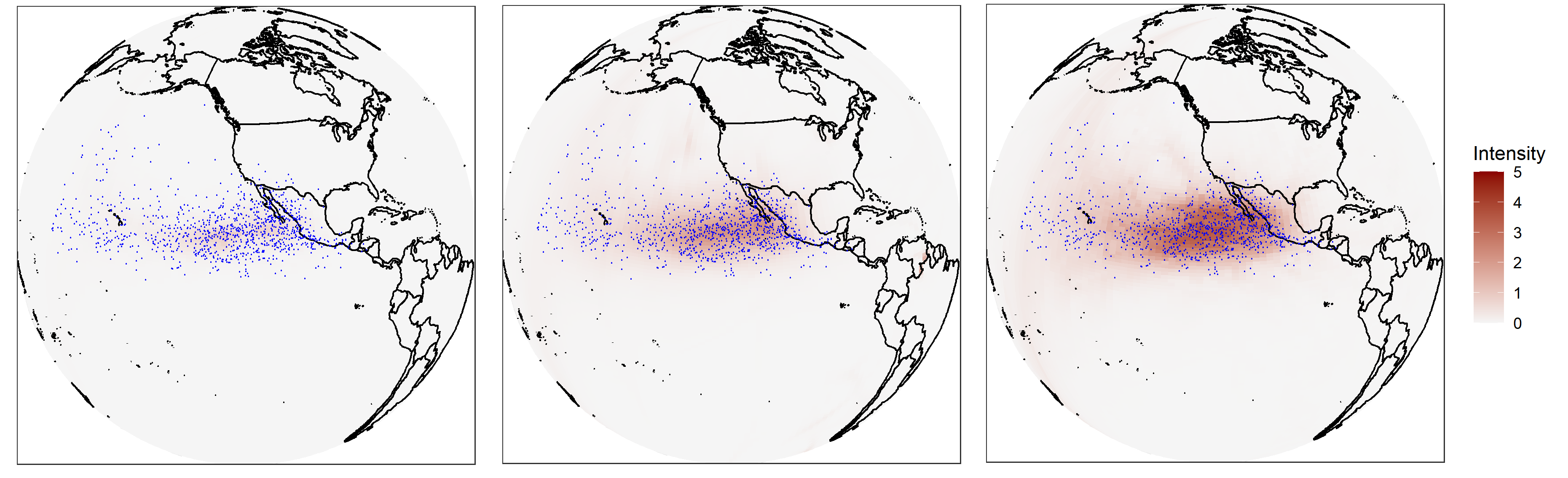}\hspace{.05em}%

  \caption{Intensity function (red surface; events per 10,000 square km) and the observed locations of cyclones (blue dots). Left panel: The empirical 10 percentile of the bootstrap distribution of the estimated intensity function based on $B = 50$ replicates. Center panel: The estimated intensity function. Right panel: Same as left panel, but for the 90 percentile.
   }
\label{pacific_plot}
\end{figure*}

\begin{figure*}[!t]

\centering

  \includegraphics[width=1.95in]{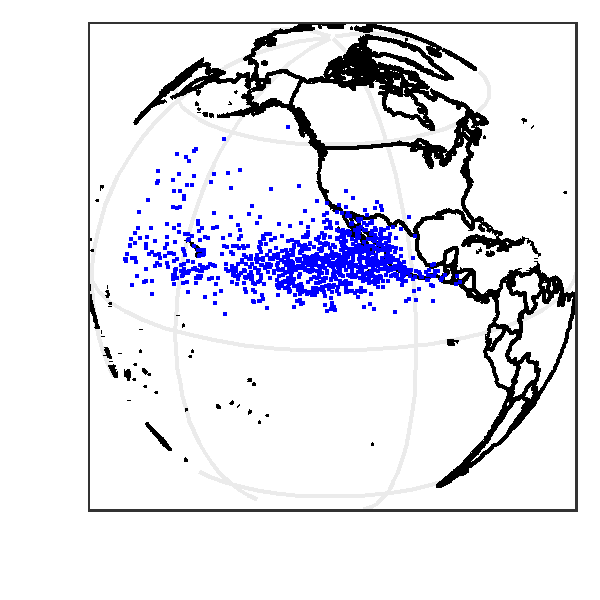}\hspace{.1em}%
  \includegraphics[width=1.95in]{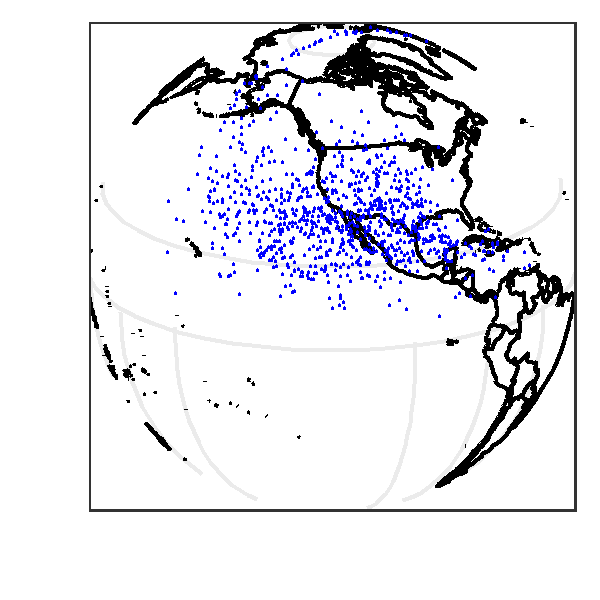}\hspace{.1em}%
  \includegraphics[width=1.95in]{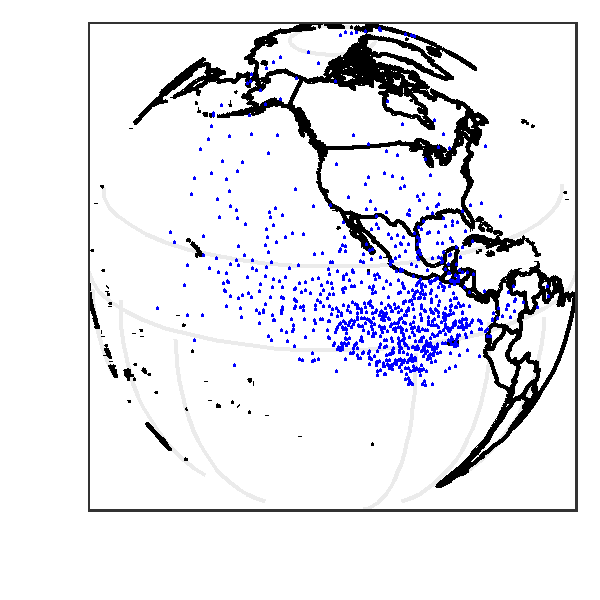}\hspace{.1em}%
  \includegraphics[width=1.95in]{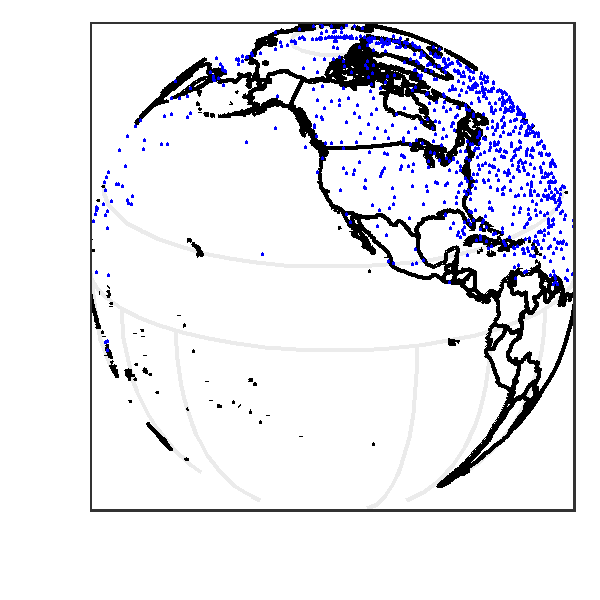}\hspace{.1em}%
\includegraphics[width=1.95in]{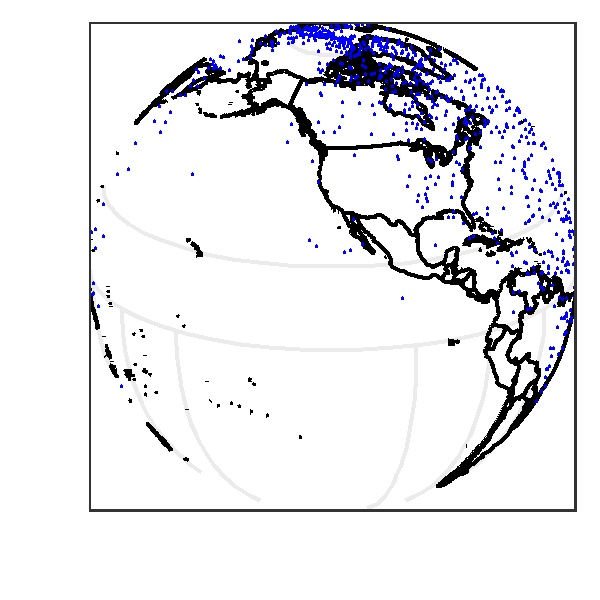}\hspace{.1em}%
\includegraphics[width=1.95in]{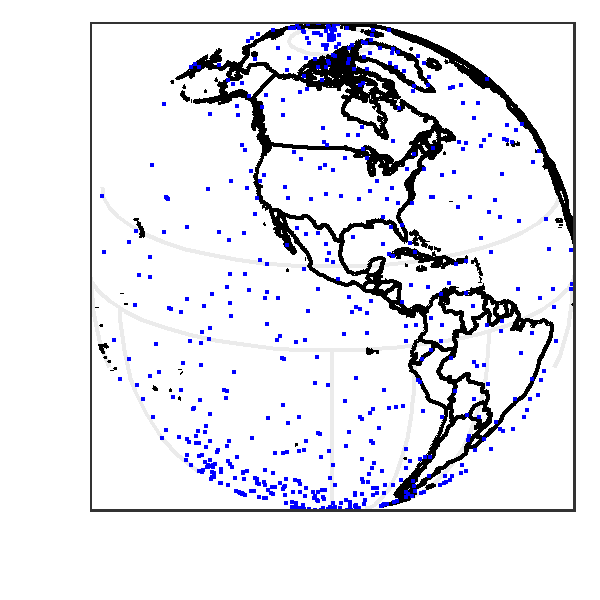}\hspace{.1em}%
\caption{Top-left panel: The end locations of the  cyclones used in this proof of concept. Top-center panel to bottom-right panel (left to right, top to bottom): The transformed locations after applying the map $G_\phi^{(k)} \circ\cdots\circ G_\phi^{(1)}(\cdot)$ to the locations, for $k = 6, 12, 18, 24, 30$, respectively. Recall that $K = 30$ for this model and that the reference density is uniform on $\mathbb{S}^2$.}
\label{pacific_flow}
\end{figure*}

In this section we apply our proposed method for spherical intensity function estimation to a dataset that contains location information on cyclones in the North Pacific Ocean. The dataset was published by the United States National Hurricane Center.\footnote{\url{https://www.nhc.noaa.gov/data/hurdat/hurdat2-nepac-1949-2020-043021a.txt}} The dataset contains six-hourly information on the location, maximum winds, and central pressure of all known tropical and subtropical cyclones (1049 in total). Cyclones are cause for significant infrastructure damage and financial loss \citep{Pielke2008}, and the modeling of cyclone prevalence is important for risk management and insurance purposes \citep[e.g.,][]{Crabbe2008}.
\\\\
Here, we focus on the end locations of the cyclones. Based on the results in Section~\ref{sec:sim}, we set the number of compositions to $K=30$ and let $p = 1$. Optimization of \eqref{obj_func3}  proved to be difficult in this case, with final estimates sensitive to the initial parameter settings. We hence used a ``committee of networks'' approach; see for example, \citet[][Section 9.6]{Bishop_1995} and \citet{Perrone_1995}. This strategy, where one trains several models with random initializations (in our case 50) and then averages their outputs, is often adopted when training highly parameterized deep networks that have multi-modal objective functions. Following model fitting, we applied the nonparametric bootstrap resampling methodology described in Section \ref{sec:mt_sphere} with $B = 50$ to obtain prediction intervals for the intensity function (without using a committee of networks in each re-fitting stage). The estimated intensity function, along with the end locations of the cyclones, is shown in the center panel of Figure \ref{pacific_plot}. The empirical 10  and 90 percentiles of the bootstrap distribution of the intensity function are shown in the left and right panels of Figure \ref{pacific_plot}, respectively.  
One can add interpretation to the inferred normalizing flow by visualizing the action of the transport maps through the transformed locations of the cyclones at different stages of the map composition. We show six of these stages in Figure \ref{pacific_flow}. Note how the transformed locations become increasingly uniformly distributed on the sphere as they are propagated through the network of compositions, as expected.
\\\\
We conclude this proof of concept with a note on computation. Fitting these highly parameterized models formed through composition is surprisingly quick and computationally efficient. Recall that the objective function \eqref{obj_func3} is simply a sum of $n$ independent quantities that can all be computed efficiently in parallel using high-dimensional arrays. Notably, no inverses are needed when evaluating the objective function or its gradient, and the model scales well with both data size $n$ and the dimension $d$. In this example, evaluating the objective function and the gradient with respect to all the parameters required approximately 0.1 s, and we therefore only needed 10 s for fitting the model with 100 gradient steps. Further, fitting for the committee of networks and bootstrapping could be trivially parallelized, so that intensity-function estimation and uncertainty quantification could in principle be done in well under a minute on a multicore platform. The high computational efficiency also makes it possible to evaluate fits from different architectures with ease. These results are encouraging, and highlight measure transport as a viable alternative to conventional LGCP modeling when, for example, the spatial point process intensity function is considerably inhomogeneous or when one is operating in a higher-dimensional space (for example, in the case of 3D-space-time point processes).

 We provide code for reproducing the results in this section at \url{https://github.com/andrewzm/SpherePP}.
 
\section{Conclusion}\label{sec:conclusion}
This paper develops a general approach to the problem of intensity-function estimation of a non-homogeneous Poisson process on the sphere. The proposed approach leverages the framework of normalizing flows by using compositions of transport maps to model the unknown intensity function. The  method is highly flexible and makes minimal assumptions on the target intensity function, which can be made arbitrary complex by increasing the number of maps in the composition and the number of basis functions in each radial flow. We illustrate the viability of our proposed approach in simulations studies, and its use in a proof of concept study where we model cyclone end locations in the North Pacific Ocean. Uncertainty quantification on the intensity function is done using  nonparametric bootstrap. 

It is not surprising that normalizing flows work well for point process intensity function estimation given their success in density estimation in several ML \& AI applications \citep[e.g.,][]{Rezende2020}. Our approach also inherits their limitations. Specifically, choosing an optimal number of maps and number of radial flows is a challenging problem. Further, optimizing the objective function \eqref{obj_func3} requires the specification of suitable initial parameter values, mini-batch size and gradient step size when doing stochastic gradient descent; these settings, which can be difficult to tune in practice, where circumvented in our case through a committee of networks approach. These difficulties are a direct consequence of choosing a highly flexible model for the intensity function, and our preliminary results indicate that this approach only begins to pay dividends when the underlying intensity function is considerably non-homogeneous. 
\\\\
Several extensions of this work are possible. First, although we have considered the nonparametric bootstrap for uncertainty quantification, one may also place prior distributions on the unknown parameters, and then implement stochastic variational Bayes as was done, for example, by \citet{Zammit_2021}. Prior distributions may be helpful in regularizing the optimization problem. Second, while we have applied the exponential map radial flows in this paper, other normalizing flows on the spheres can also be considered, which might be more suitable for modeling point processes typically encountered in spatial statistics. Third, the incorporation of covariate information will be important for many practical applications, and an extension of our measure transport approach that allows for covariates is an interesting avenue of future research. Fourth, while we consider the setting of inhomogeneous Poisson process on $\mathbb{S}^{2}$ this work, the general normalizing flows framework is applicable to higher dimensional spheres. In higher dimensional settings, the normalizing flow approach will have significant computational advantages over LGCPs.  Finally, theoretical properties of normalizing flows on Euclidean domains have been investigated where it has been shown that under mild conditions arbitrary probability distributions and intensity functions can be approximated arbitrarily well using normalizing flows with relatively simple transformations \citep{Huang2018, Jaini2019, Ng2020}. It is desirable to investigate whether normalizing flows on spherical and other non-Euclidean domains share such nice properties.

\section*{Acknowledgements}

Andrew Zammit-Mangion's research was supported by the Australian Research Council Discovery Early Career Research Award (DECRA) DE180100203. The authors would like to thank Yi Cao and Bao Vu for discussions related to the source code and the manuscript.

\section*{Declarations of Interest}
Declarations of interest: none.

\bibliographystyle{asa} 
\bibliography{refs}





\end{document}